\documentclass[prd,aps,preprint,tightenlines,showpacs,nofootinbib,superscriptaddress]{revtex4-1}
\usepackage{mathrsfs}
\usepackage{amsfonts}
\usepackage{amsmath}
\usepackage{array}
\usepackage{verbatim}
\usepackage{bm}
\usepackage{epsfig}
\usepackage{graphicx}
\usepackage{relsize}
\RequirePackage{xspace}

\begin{document}
%\selectlanguage{english}
\title{ Study of gluon GPDs in exclusive $J/\psi$ production in electron-proton scattering}

\author{\firstname{S.~V.}~\surname{Goloskokov}}
\email[]{goloskkv@theor.jinr.ru}
\affiliation{
Bogoliubov Laboratory of Theoretical Physics, Joint Institute for Nuclear Research, Dubna 141980, Moscow region, Russia}
\affiliation{Institute of Modern Physics, Chinese Academy of Sciences, Lanzhou 730000, China}

\author{\firstname{Ya-Ping}~\surname{Xie}}
\email[]{xieyaping@impcas.ac.cn}
\affiliation{Institute of Modern Physics, Chinese Academy of Sciences, Lanzhou 730000, China}
\affiliation{University of Chinese Academy of Sciences, Beijing 100049, China}

\author{\firstname{Xurong}~\surname{Chen}}
\email[]{xchen@impcas.ac.cn}
\affiliation{Institute of Modern Physics, Chinese Academy of Sciences, Lanzhou 730000, China}
\affiliation{University of Chinese Academy of Sciences, Beijing 100049, China}
\affiliation{South China Normal University, Guangzhou 510006, China}

\begin{abstract}
Exclusive photoproduction of heavy $J/\psi$ meson is performed within handbag factorization method. The General Parton Distributions (GPDs) for gluons which are an essential ingredient here, are constructed using double distribution (DD) representation.
The calculated cross section for few sets of gluon GPDs $H_g$ describe fine $J/\psi$ production experimental data from
HERA to LHC energies. Based on this approach we estimate $J/\psi$ spin asymmetries determined by $E_g$ and $\tilde{H}_g$.
The GPDs factorization approach can be employed to analyse heavy vector meson production in electron-proton scattering.
Investigations of this reactions at future electron-ion colliders in USA and China can give an essential information on gluon GPDs.
\end{abstract}
\maketitle

\section{Introduction}\label{intro}
Exclusive heavy meson production (EHMP) is important to study the nucleon structure in high energy regions. It can be employed to
study the property of the gluon density of proton. In electron-proton scattering and ultraperipheral heavy ions collisions, photoproduction
of heavy meson  is an important probe to study the nucleon structure.

EHMP was studied for the first time in pQCD \cite{ZPhysC57} where it was shown that the process amplitude is proportional to gluon density. This method, based on the leading logarithmic approximation was extended to the light meson production \cite{Br94}, where amplitude factorized into hard scattering part and gluon density in proton. This property was confirmed later on in next to leading order approximation \cite{EPJC34}.

Several methods have been employed to study EHMP in QCD. One of the popular approach is based on the $k_T$-factorization method \cite{PPN37}, where the process amplitude is determines as a convolution of unintegrated $k_T$- dependent gluon distribution and a hard part obtained by resummation of large $\ln{1/x}$ terms.
 In dipole model where factorization was shown as well \cite{PRD68}, the colour dipole amplitude formed between initial photon and final vector meson, interacts with proton via the gluon exchange. This model was used e.g. in \cite{PRD74,PRD78} to study vector meson production processes.

 Unfortunately, in all these models, it is problematic to consider effects of skewness which is important in exclusive processes. Such information contains in GPDs, which are an essential ingredients of exclusive reactions.
GPDs can be studied in Deeply Virtual Compton Scattering \cite{ji1,rad,dvcs}, Deeply Virtual Meson Production processes
\cite{GPD-review01,GPD-review02, dvmp2}, Time-like Compton Scattering \cite{Berger:2001xd,  Xie:2022vvl} and Neutrino Nucleon processes\cite{Pire:2017tvv}. It was shown that these processes factorized \cite{ji1,rad,dvcs} into the hard subprocess and GPDs that depend on three variables $x$-the longitudinal fraction of proton momentum, $\xi$, skewness - longitudinal momentum transfer and $t$- momentum transfer. In the forward limit ($\xi, t=0$) GPDs are equal to PDF, and gives
information on the longitudinal hadron structure. Moments of GPDs are connected with the hadron form-factors (FFs) \cite{dvmp2} and give information on hadron  distributions in the transverse plane.  Thus GPDs give us information on
the 3D structure of nucleons \cite{3d}. GPDs can analize the nucleon spin property via the Ji's sum rules\cite{ji1}. In DVCS and light vector meson production, the quark GPDs and gluon GPDs contribute to the cross sections. More information on GPDs can be found in reviews \cite{GPD-review01, GPD-review02}.
The heavy vector mesons like $J/\psi, \Upsilon$ are constructed by the $c$ or $b$ quarks, which are practically invisible in proton. As a result, the quark GPDs can be neglected in EHMP. Thus, these reactions are an unique  processes to study the gluon GPDs.

There are a number of dynamical models which were used to construct GPDs \cite{din_gpds}. The popular models were based on the double distribution representation(DD) \cite{mus99} where used to study DVCS \cite{dvmp_m} and  DVMP \cite{GPD-review01, dvmp2, gk06}. The last one - GK model was successfully used to study light vector meson
production \cite{gk06, gk08} and pseudoscalar meson production \cite{gk09}. These results were extended to study pseudoscalar meson production at lower energies \cite{Goloskokov:2022mdn}.

Electron-Ions Collider (EIC) is an important platform to investigate nucleon GPDs. It can help us to study nucleon structure and 3-Dimenson parton distributions. The USA and China both plan to build EICs\cite{eic, eicc}. In USA EIC and EIC of China (EicC), $J/\psi$ cross sections and spin asymmetries will be measured to study GPD properties.

\begin{figure}[!h]
	\centering
	\includegraphics[width=12cm]{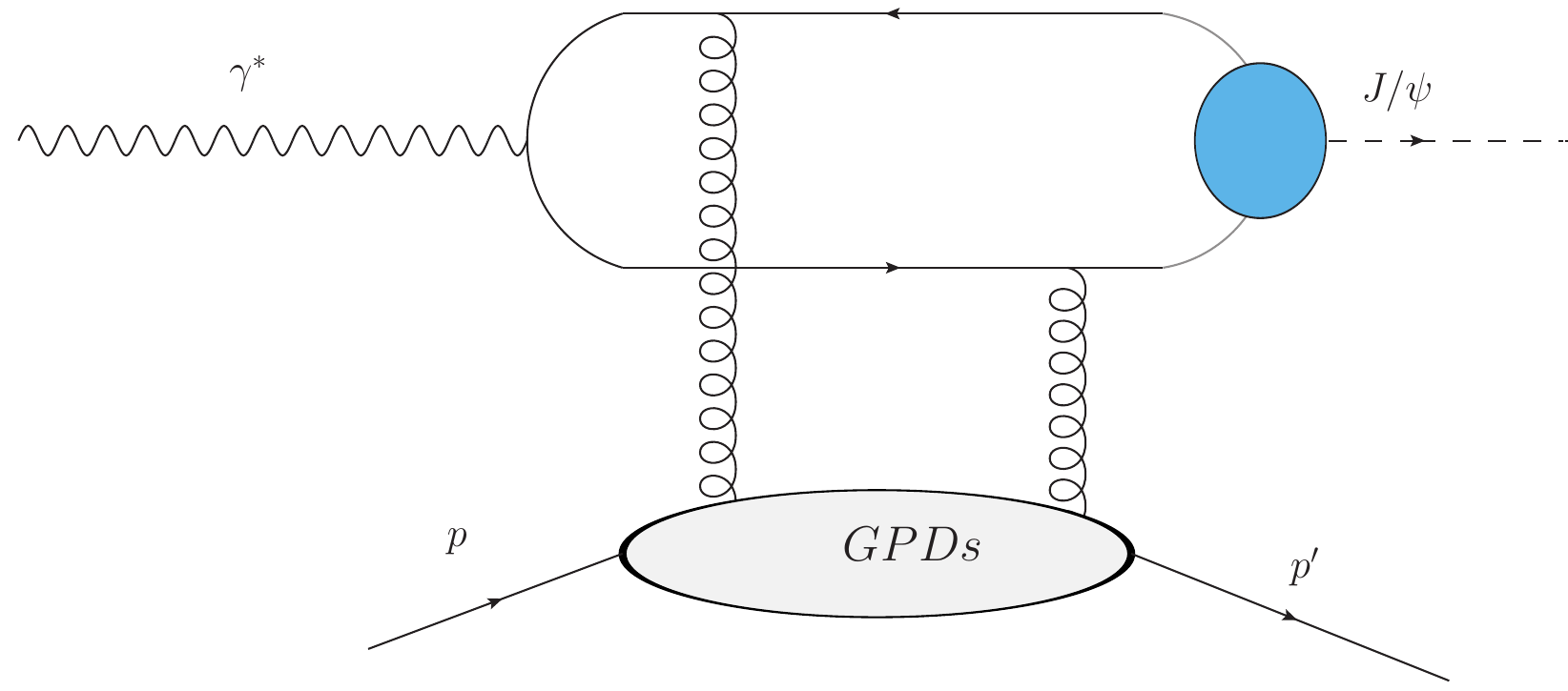}
	\caption{Typical diagram of EHMP  in photon-proton scattering. }
	\label{figvm}
\end{figure}

In EHMP, the gluon GPDs is important In Fig.~\ref{figvm}, the typical diagram for heavy vector meson production \cite{gk09} is depicted. The process factorized into hard part with heavy quark loop, that contain nonpertubative vector meson wave function and gluon GPDs. The gluon GPD functions can be employed in the amplitude of the heavy vector meson production process. The hard part can be calculated perturbatively. Moreover, large meson mass in propagators generates the hard scale $\propto m_v$  in a process that saves factorization properties even for photoproduction
$Q^2$ =0 $\mbox{GeV}^2$ \cite{ZPhysC57}.

This paper consists of the following sections.
In Sec.II, the heavy vector meson differential cross section formalism is exhibited
with a brief introduction to the GPDs. In Sect. III we discuss the spin observables. In Sec. IV, we discuss the model for GPDs.
In Sec.V, results of our model calculations are compared with experimental data.
Some discussion and conclusions are presented in Sec.VI.
\section{Theoretical frame and model}
In exclusive vector meson production process, the virtual photon scatters off proton and produce heavy vector meson. The process is depicted in Fig.~\ref{figvm},
In this process $\gamma^*(q)+P(p)\to V(q')+P(p')$,
we have following variables
\begin{equation}
Q^2=-q^2, t=(p-p')^2, W^2=(p+q)^2, m_V^2=q'^2.
\end{equation}
We can use the Bjorken's variable and skewness parameters for heavy vector meson as
\begin{equation}
x_B=\frac{m_V^2+Q^2}{W^2+Q^2},\;\; \xi=\frac{x_B}{2-x_B}.
\end{equation}
From Fig.~\ref{figvm}, it can be seen that the gluons interact with the quark and antiquark and the GPDs factorization method
can be used in heavy vector meson production process.   	
The gluon GPDs
$H^g$, $E^g$, $\widetilde{H}^g$ and $\widetilde{E}^g$ are related with matrix elements of gluon field operators.
\begin{eqnarray}
	\langle p'\nu'|\sum_{a,a'} A^{a\rho}(0)\, A^{a'\rho'}(\bar{z})|p\nu\rangle
	&=& \frac{1}{2}  \sum_{\lambda=\pm 1} \epsilon^\rho(k_g,\lambda)\,
	\epsilon^{*\rho^\prime} (k_g,\lambda') \\
&\times& \int_0^1
\frac{dx}{(x+\xi- i\epsilon)(x-\xi + i\epsilon)}
{\rm e}^{-i(x-\xi) p\cdot\bar{z}} \nonumber\\
&\times& \left\{\frac{\bar{u}(p'\nu')\,n\!\!\!\slash\, u(p\nu)}{2\bar{p}\cdot
	n}\;H^g(x,\xi,t)
+ \frac{\bar{u}(p'\nu')\,i\, \sigma^{\alpha\beta}\,n_\alpha
	\Delta_\beta\, u(p\nu)}{4m\, \bar{p}\cdot n}\;{E}^g(x,\xi,t)\right.\nonumber\\
&+& \left.\lambda \frac{\bar{u}(p'\nu')\,n\!\!\!\slash\gamma_5\, u(p\nu)}{2\bar{p}\cdot
	n}\;\widetilde{H}^g(x,\xi,t)
+ \lambda \frac{\bar{u}(p'\nu')\,n\cdot\Delta\,\gamma_5\,
	u(p\nu)}{4m\, \bar{p}\cdot n}\;\widetilde{E}^g(x,\xi,t)\right\}\,.\nonumber
\label{eq:matrix}
\end{eqnarray}

The gluon contribution to light vector meson electroproduction within GPD approach were calculated in \cite{gk06}. The apprroach can
be employed to study heavy vector meson production where only gluons contributions are taken into account in the amplitudes.
The helicity conservation amplitude of heavy vector meson production is given as
\begin{eqnarray}
\mathcal{M}_{\mu^\prime+,\mu+}
&=& \frac{e}{2}C_V\int_0^1\frac{dx}{(x+\xi)(x-\xi+i\epsilon)}\nonumber\\
&&\times\big\{\mathcal{H}^{V+}_{\mu^\prime,\mu}\,H_g(x, \xi, t,\mu_F)   +\mathcal{H}^{V-}_{\mu^\prime,\mu}\,\tilde{H}_g(x, \xi, t, \mu_F]\big\}.
 \end{eqnarray}
While the helicity flip amplitude can be written as
\begin{eqnarray}
\mathcal{M}_{\mu^\prime-,\mu+}
&=&-\frac{e}{2}C_V\frac{\sqrt{-t}}{2m}\int_0^1\frac{dx}{(x+\xi)(x-\xi+i\epsilon)}\nonumber\\
&&\times\big\{\mathcal{H}^{V+}_{\mu^\prime,\mu}\,E_g(x, \xi, t,\mu_F)
+\mathcal{H}^{V-}_{\mu^\prime,\mu}\,\tilde{E}_g(x, \xi, t,\mu_F)\big\}.
\end{eqnarray}

Here the amplitudes $\mathcal{H}^{V\pm}_{\mu^\prime,\mu}$ are determined as a sum and differences of amplitudes with different gluon helicities.
\begin{equation}
\mathcal{H}^{V\pm}_{\mu^\prime,\mu} = \big[\mathcal{H}^V_{\mu^\prime+,\mu+}\pm\mathcal{H}^V_{\mu^\prime-,\mu-}\big],
\end{equation}
and flavor factor $C_V= C_{J/\psi}=2/3$.

In Ref.\cite{gk06}, the hard scattering amplitude was calculated by considering the transverse momenta in the quark
loop (Fig.~\ref{figvm}) with taken into account the gluon radiation effects. Resummation of theses effects \cite{sterman} provides the Sudakov fact that suppress large quark-antiguark separations. Such
contributions appear when quark or antiquark in the quark loop have small momenta.
In what follows we discuss nonrelativistic vector meson wave function, when the quark
momenta are equal each other. As a result, the Sudakov effects in heavy vector meson
production should be rather small and can be omitted  here. The quark transverse
momenta are taken into account.

To calculate hard scattering amplitude we consider six gluon Feynman diagrams which are presented in Ref.\cite{EPJC34,gk06} . After a length calculations,
the hard amplitude can be cast into
\begin{eqnarray}
\mathcal{H}^{V\pm}_{\mu^\prime,\mu}(x, \xi)=64\pi^2\alpha_s(\mu_R)\int_0^1d\tau
\int \frac{d^2\mathbf{k}_\perp}{16\pi^3}\psi(\tau, \mathbf{k}_\perp)\mathcal{F}_{\mu^\prime,\mu}^{\pm}(\tau, x, \xi, \mathbf{k}^2_\perp).
\end{eqnarray}
Here $\tau$ and $1-\tau$ are the fraction of longitudinal part of quark (antiquark) momenta incoming to the meson wave function, $\mathbf{k}_\perp$ is there transverse part.
The $k$ -dependent wave function of the vector meson is written as
\begin{equation}
\psi(\tau, \mathbf{k}_\perp)=a_{v}^2f_{v}\exp\Big(-a_{v}^2\frac{ \mathbf{k}^2_\perp}{\tau(1-\tau)}\Big).
\end{equation}
Here $f_{v}$ is a  $J/\psi$ decay constant, the parameter $a_{v}$  is fixed from the best fit $J/\psi$ cross section and determine the average value of $\langle\mathbf{k}\rangle ^2_\perp$.

Note, that the full result for amplitudes for arbitrary $\tau$ are rather complicated.
In what follows we simplified our results using nonrelativistic form of wave function when quark and antiquark have the same momenta, $\tau=1/2$. This means that we use $\delta(\tau-1/2)$ in Eq.~(8), which is typical
for most  calculations see e.g. \cite{ZPhysC57}.
For $\tau = 1/2$, the hard part of the amplitude can be written as
\begin{eqnarray}
\mathcal{F}_{\mu^\prime,\mu}^{\pm}=\frac{f^{\pm}_{\mu^\prime,\mu}}{(2\mathbf{k}^2_\perp+m_V^2+Q^2)(4\xi\mathbf{k}_\perp^2+(m_V^2+Q^2)(\xi-x)+i\epsilon
	)(4\xi\mathbf{k}_\perp^2+(m_V^2+Q^2)(\xi+x))}
\end{eqnarray}
It can be seen that the typical variable that appears in quark propagators is $\mu^2=m_V^2+Q^2$. For $J/\psi$ production the  scale $\mu^2 \sim$ 10 $\mbox{GeV}^2$ for $Q^2=0$ and factorization form can be used even for photoproduction case.
We define renormalization scale $\mu_R=\mbox{max}(\tau\, \mu,(1-\tau)\, \mu)=\mu/2$, based on \cite{gk06} and $\mu_F=\mu$ \cite{Koempel:2011rc}. Note, that within GPDs approach EHVM was investigated e.g. in \cite{Koempel:2011rc}. However, mass terms in propagators of hard scattering amplitude are not taken into account there.

For longitudinal and transverse helicity conservation amplitudes $f^{+}_{\mu,\mu}$ have a form
\begin{eqnarray}
f^{+}_{00}&=&-64\sqrt{Q^2}(m_V^2+Q^2)^2(x^2-\xi^2),\\
f^{+}_{11}&=&64m_V(m_V^2+Q^2)^2(x^2-\xi^2).
\end{eqnarray}
Here  we omit $k$ dependent terms.
For the $f^{-}_{\mu,\mu}$ than contains $\tilde {H}$ contribution, we find that
 \begin{equation}
f^{-}_{11}=- 256 (m_V^2+Q^2) \mathbf{k}_\perp^2\, m_V\, x\, \xi,\;\; f^{-}_{00}=0.
 \end{equation}
 From Eq.~(10) and (11),  it can be found the relations between longitudinal and transverse amplitudes can be written as \cite{ZPhysC57}
  \begin{equation}
\mathcal{M}_{0\nu',0\nu}=-\frac{\sqrt{Q^2}}{m_V} \mathcal{M}_{+\nu',+\nu}.
 \end{equation}
 It is valid for the case of the same form of the meson wave function  Eq.~(8) for longitudinally and transversally polarized vector mesons.
 In this work, we use this approximation for simplicity.

\section{Spin asymmetry observables}
In heavy vector meson production, some asymmetries can be adopted to constrain the GPD functions.
Initial state helicity correlations $A_{LL}$ can be measured with longitudinally polarized beam and target. It can be expressed in terms of helicity amplitudes as \cite{gk06}
\begin{eqnarray}
A_{LL}(ep\to epV) = 2\sqrt{1-\epsilon^2}\frac{\mathrm{Re}[\mathcal{M}^H_{++,++}\mathcal{M}^{\tilde{H}*}_{++,++}]}
{\epsilon|\mathcal{M}^H_{0+,0+}|^2+|\mathcal{M}^H_{++,++}|^2}.
\end{eqnarray}
It can be seen that $A_{LL}$ can be used to study property of polarized GPDs $\tilde{H}$ of gluon.

On the other hand, the single spin asymmetry $A_N$ can be also measured at heavy vector meson production.
For the photoproduction case when the longitudinal amplitude  can be omitted the $A_N$ asymmetry can be written as \cite{Koempel:2011rc}
\begin{eqnarray}
A_N(ep\to epV) =-\frac{2\mathrm{Im}[\mathcal{M}^H_{++,++}\mathcal{M}^{E*}_{+-,++}]}{|\mathcal{M}^H_{++,++}|^2+|\mathcal{M}^E_{+-,++}|^2}.
\end{eqnarray}

The double single spin asymmetry $A_{LS}$ can be obtained via amplitudes \cite{Koempel:2011rc}
\begin{eqnarray}
A_{LS}(ep\to epV)=\frac{2\mathrm{Re}[\mathcal{M}^H_{++,++}\mathcal{M}^{E*}_{+-,++}]}{|\mathcal{M}^H_{++,++}|^2+|\mathcal{M}^E_{+-,++}|^2}.
\end{eqnarray}
Thus, both single spin $A_N$ and double spin $A_{LS}$ asymmetries can be employed to study $E_g$ GPD function since $M_{+-,++}$ amplitude
 is expressed via $E_g$ GPDs. If the asymmetries $A_N$ and $A_{LS}$
are measured at future, they can be applied to constrain $E_g$ GPD functions.
\section{GPD model}
The skewness $\xi$ dependencies of GPDs can be reconstructed with the help of DD representation \cite{mus99} which connect GPDs with PDFs.
\begin{equation}
F_i(x, \xi, t) = \int_{-1}^1d\beta \int_{-1+|\beta|}^{1-|\beta|}d\alpha \, \delta(\beta + \xi \alpha -x)f_i(\beta, \alpha, t).
\end{equation}
where $F_i = H$, $\tilde{H}$, $E$, $\tilde{E}$ GPDs and the double distribution functions $f_i$ for gluon contribution are given by the following expressions
\begin{eqnarray}
f_i(\beta, \alpha, t) = e^{B_V t}\,h_i(\beta,\mu_F) \frac{15}{16}\frac{[(1-|\beta|)^2-\alpha^2]^2}{(1-|\beta|)^{5}}.
\end{eqnarray}
Here $h_i(\beta,\mu_F)$ is a corresponding PDFs at the scale $\mu_F$, $B_V$ is a slope parameter. We use  for $B_V$ a little bit modified form from \cite{gk06,gk07}
\begin{equation}
B_V= b_0+b_1 \ln[m_V^2/(m_V^2+Q^2)]+\alpha' \ln[(W^2+Q^2)/(m_V^2+Q^2)].
\end{equation}
Here $\alpha'$=0.185 (GeV$^{-2})$ is a slope of Pomeron trajectory determiner by DESY experiments \cite{Levy:2007pub}.
Parameters $b_0$=1.06 $(\mbox{GeV}^{-2})$ and $b_1$=0.13 $(\mbox{GeV}^{-2})$ are found from the best fit $W$ and $t$ dependencies of $J/\psi$ production cross section at DESY energies. For the cross section we have $d \sigma/dt\sim \exp (2 B_V t)$. For example, photoproduction case we find $2 B_V(W$=40 $\mbox{GeV})$=4.01 $( \mbox{GeV}^{-2})$ and $2 B_V(W$=100 $\mbox{GeV})$=4.69 $(\mbox{GeV}^{-2})$, which is closed to H1 result \cite{EPJC24}.

In our analysis to determine GPDs we test three sets of PDFs\cite{Alekhin:2018pai,Hou:2019qau,NNPDF:2014otw} . We generate here $Q^2$ evolution of GPDs through
evolution of PDFs. It was found in \cite{gk08} that this approximation is close to $H^g$ evolution obtained by Vinnikov code \cite{Vinnikov:2006xw} at least till  $Q^2=$ 40 $\mbox{GeV}^2$. Nevertheless to avoid problems with evolution we fit PDFs
at the scale $\mu_0^2$=10 $\mbox{GeV}^2$ which is closed to $\mu^2$ for $Q^2$=0 $\mbox{GeV}^2$. Evolution for $Q^2$>0 $\mbox{GeV}^2$ is done with the help of PDFs evolution.

\section{Numerical results and discussion}
In this section we compare our model results with existing experimental data of $J/\psi$ production at HERA and
perform some prediction for future accelerators.
 The longitudinal and transversal differential cross sections as functions of $|t|$ and total cross
sections of heavy vector meson in photon-proton scattering as function of W and Q$^2$ are calculated as
\begin{eqnarray}
&&\frac{d\sigma_T}{dt}=\frac{1}{16\pi W^2(W^2+Q^2) }\big[|\mathcal{M}_{++,++}|^2+|\mathcal{M}_{+-,++}|^2\big],\;\quad \frac{d\sigma_L}{dt}=\frac{Q^2}{m_V^2} \frac{d\sigma_T}{dt};\\
&&\frac{d\sigma}{dt}= \frac{d\sigma_T}{dt}+\epsilon \frac{d\sigma_L}{dt}.
\end{eqnarray}
Here $\epsilon$ is a polarization of the virtual photon which in our case is close to unity.
As mentioned before, result for longitudinal cross section,  Eq.~(21)  is valid for the case of the same form of the wave function for different
vector meson polarization. We use this approximation in this work.

The total, integrate over $|t|$ cross section is estimated as
\begin{equation}
\sigma=\int_{t_{min}}^{t_{max}}\frac{d\sigma}{dt}\simeq \frac{1}{2B_V}\frac{d\sigma}{dt}(t=0).
\end{equation}
With the above calculation, we can obtain the differential cross sections of vector meson as a function of $|t|$ and integrated over $|t|$ cross sections
as a function of W and Q$^2$.

\begin{figure}[!h]
	\centering
		\includegraphics[width=7.5cm]{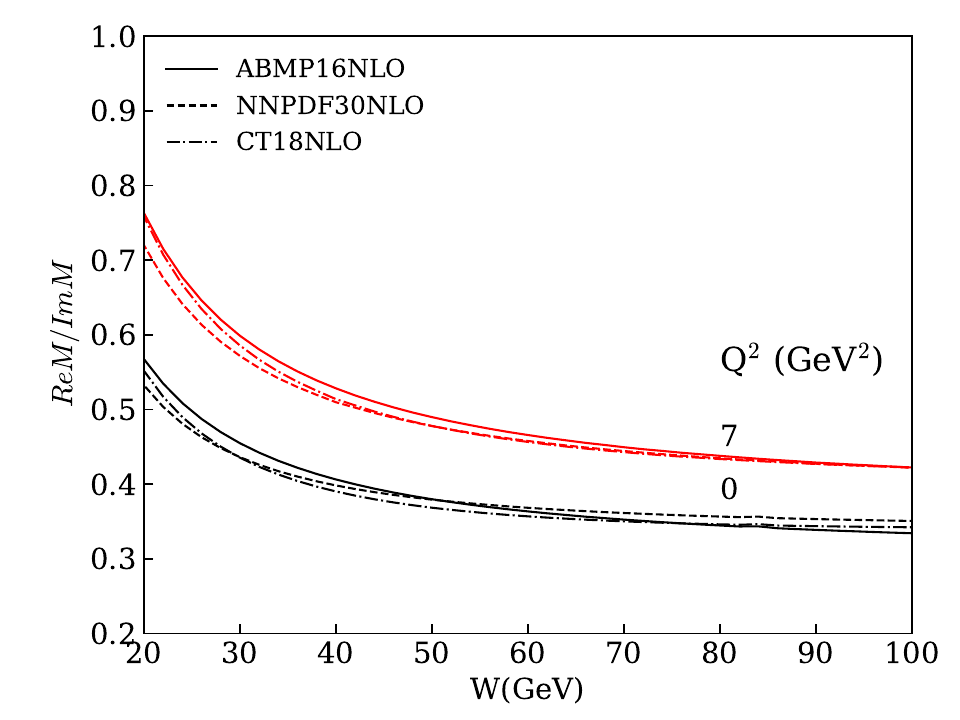}
	\includegraphics[width=7.5cm]{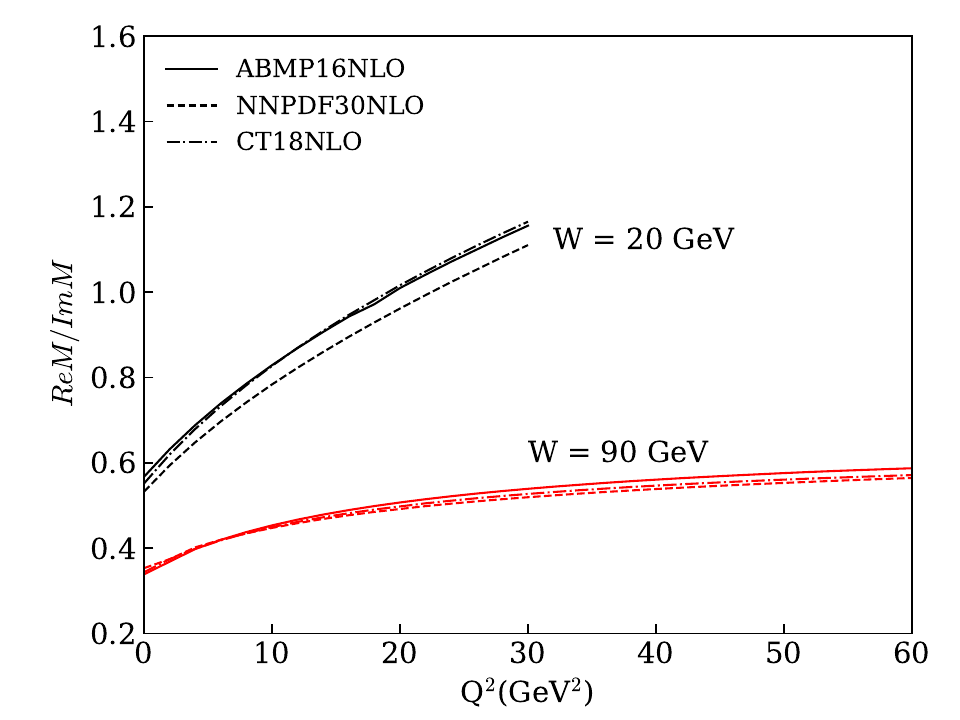}
	\caption{Ratio of  $ReM/Im M$ parts of $J/\psi$ amplitudes at fixed W vs Q$^2$. }
	\label{fig00a}
\end{figure}

 In this work, three models of gluon density  ABMP16NLO\cite{Alekhin:2018pai}, CT18NLO \cite{Hou:2019qau} and NNPDF30NLO \cite{NNPDF:2014otw} are adopted 
  in the calculations of GPDs. 
The $a_V$ parameter  determined the average value of $k_\perp^2$ in propagators of the hard amplitude Eq.~(9).  We use here
$a_V^2$=0.091 $\mbox{GeV}^{-2}$ that gives $\langle k_\perp^2\rangle \sim$ 1 $\mbox{GeV}^{2}$ and fits experimental data well.
Model permits us simultaneously calculate real and imaginary parts of production amplitude (see Eq.~(9)). Our predictions for $\mathrm{R}=ReM/Im M$ ratio are presented on
 Fig.~\ref{fig00a} as a function of W for two values of $Q^2$ and as a function of
$Q^2$ for two typical energies $W =20$ $ \mbox{GeV}$ at EicC and $W=90$ $ \mbox{GeV}$ at EIC-US.
 We find that R-ratio at fixed $Q^2$ decreases with energy and it grows with
 $Q^2$ at fixed W. Moreover, at EicC energy range, R- ratio is not far from
 unity. At higher EIC-US energy, ratio is near  0.5.

\begin{figure}[!h]
	\centering
	\includegraphics[width=7.5cm]{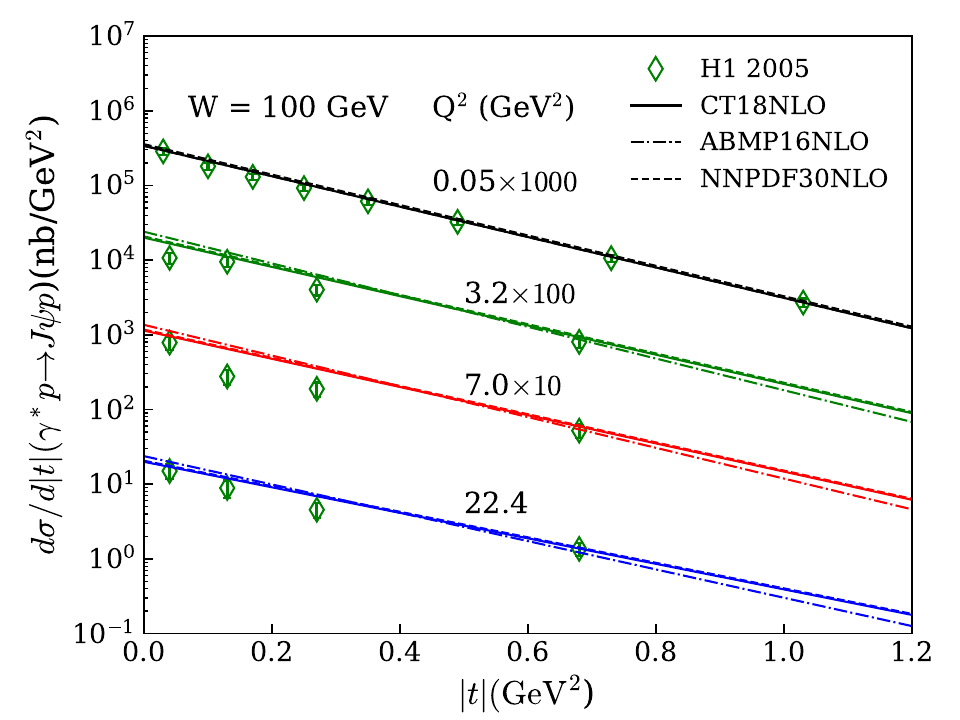}
	\includegraphics[width=7.5cm]{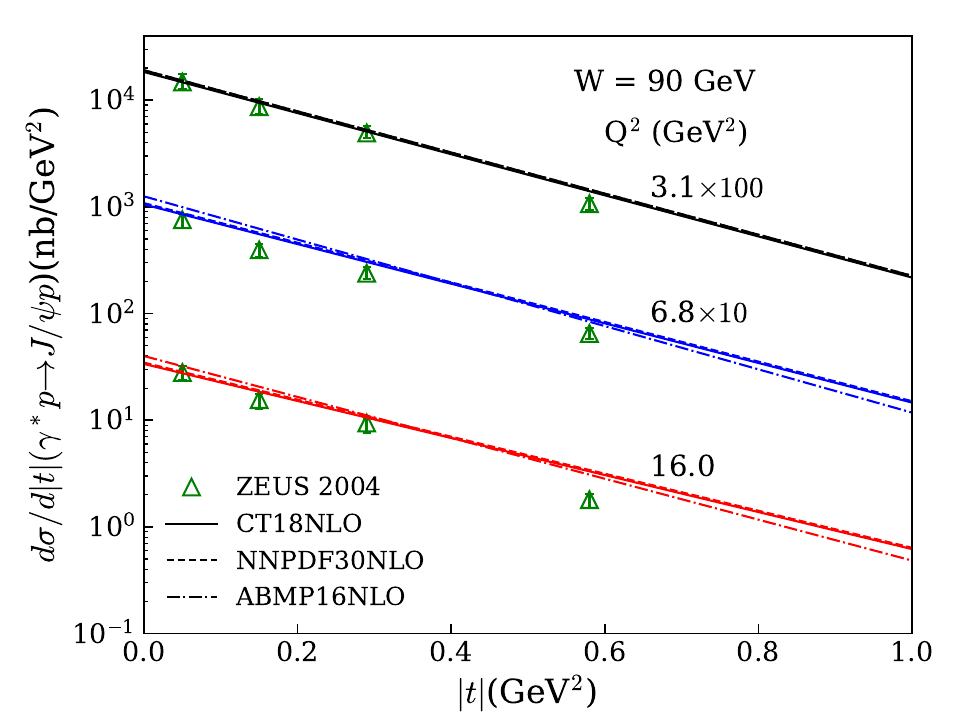}
	\caption{$J/\psi$  differential cross section vs $|t|$ at fixed W and different $Q^2$.  The HERA experimental data are from
	Ref. \cite{NPB695} and \cite{EPJC46}.Cross sections are scaled by the factor shown in the graph. }
	\label{fig00}
\end{figure}

\begin{figure}[!h]
	\centering
	\includegraphics[width=7.5cm]{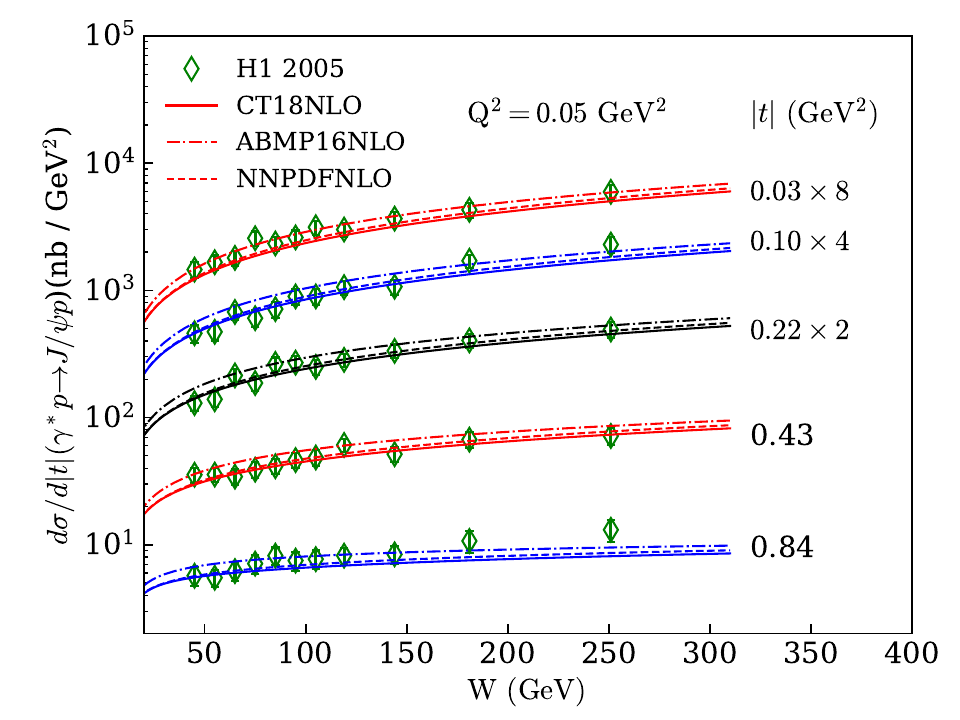}
	\includegraphics[width=7.5cm]{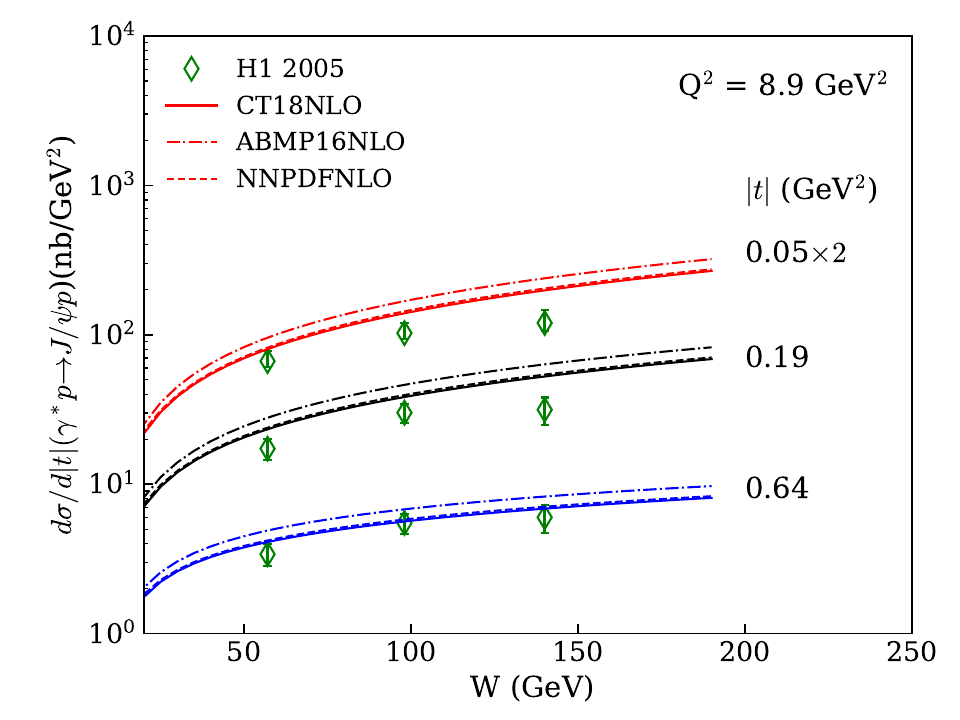}
	\caption{ $J/\psi$  differential cross section as a function of W at different $|t|$. The H1 experimental data
	are from Ref.\cite{EPJC46}. }
	\label{fig02}
\end{figure}

Firstly, we compare our prediction of three GPD models based on PDFs (ABMP, NNPDF, CT18) \cite{Alekhin:2018pai,Hou:2019qau,NNPDF:2014otw}  with
available experimental data on the differential cross sections of $J/\psi$ production from DESY  ZEUS \cite{NPB695} and H1 \cite{EPJC46} experiments
that was done at HERA.
In Fig.~\ref{fig00}, the  $|t|$ dependence of differential cross sections of $J/\psi$ at fixed W and $Q^2$ are exhibited. It can be seen that all models give good
descriptions to the differential cross sections. We can conclude that our choice of $B_V$ slope in Eq.~(20) which is essential in description of $|t|$ -dependencies
of cross section is quite good.
In Fig.
\ref{fig02}, our results for differential cross sections of $J/\psi$ together with H1 experimental data from Ref.\cite{EPJC46} are depicted.
In these two graphs, we show W-dependence of the differential cross sections  for two $Q^2$ values and different $|t|$. All models describe the
 differential cross sections at different $|t|$ quite well. This indicates that W-dependence $B_V$ slope in Eq.~(20) is a good choice.

\begin{figure}[!h]
	\centering
	\includegraphics[width=7.5cm]{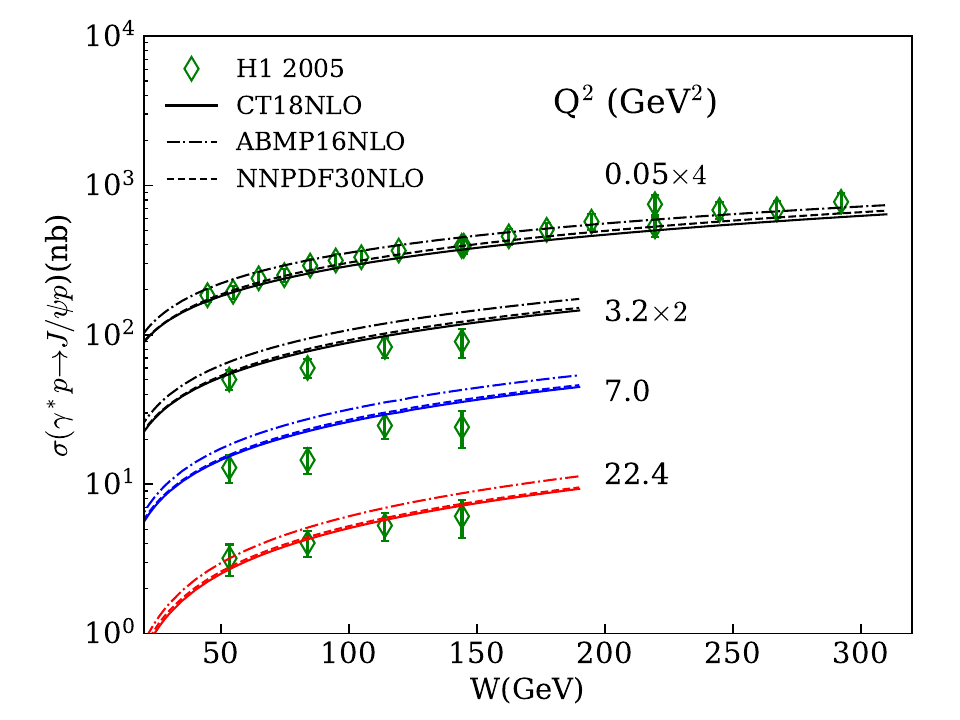}
	\includegraphics[width=7.5cm]{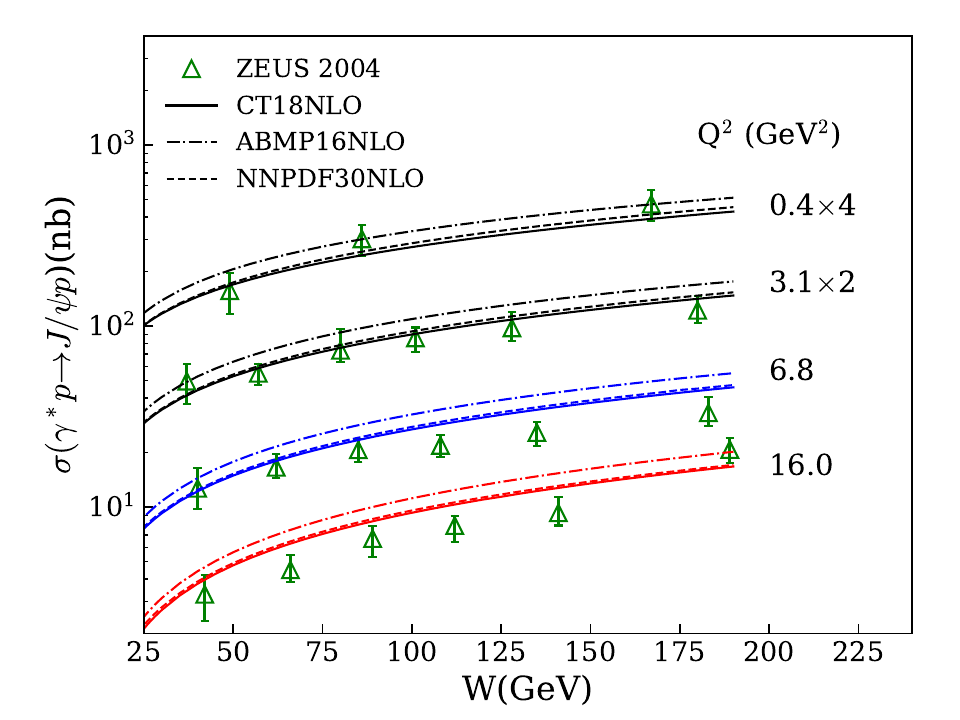}
	\caption{$J/\psi$ total cross section vs W at different $Q^2$ comparing with the HERA experimental data
	are taken from Ref. \cite{NPB695} and \cite{EPJC46}.}
	\label{fig03}
\end{figure}

Moreover, we show our prediction of total cross sections of $J/\psi$ compared with H1 and ZEUS experiments data measured at DESY.
 In Fig.~\ref{fig03} they are presented  as a function of $W$ at different $Q^2$. It can be found that variants of GPDs based on all investigated
 models give a good agreement to the experimental data.

\begin{figure}[!h]
	\centering
	\includegraphics[width=7.5cm]{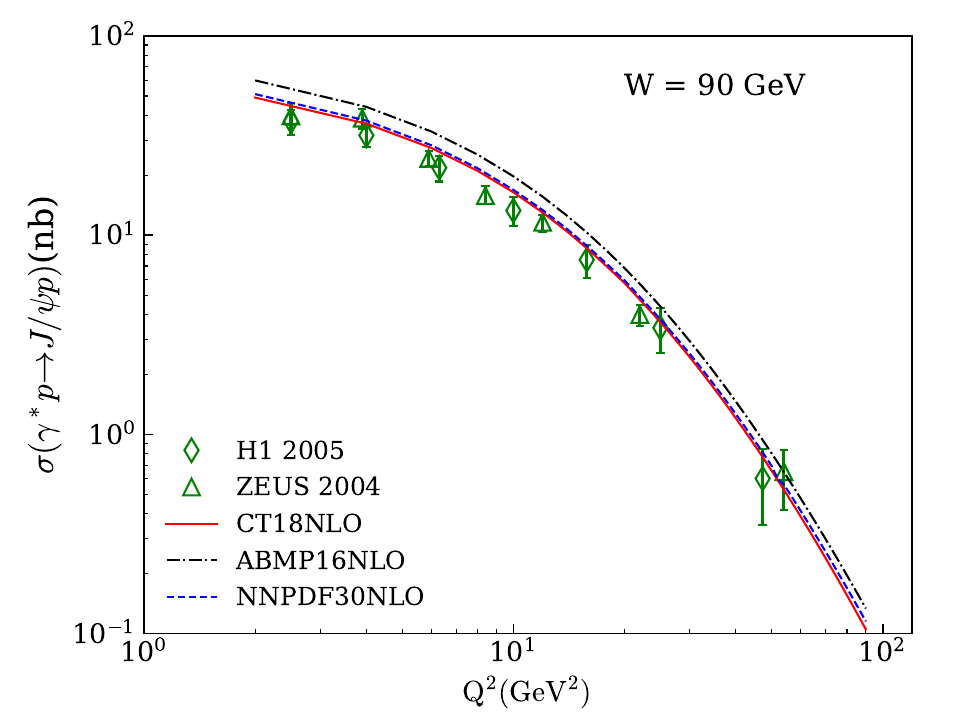}
	\includegraphics[width=7.5cm]{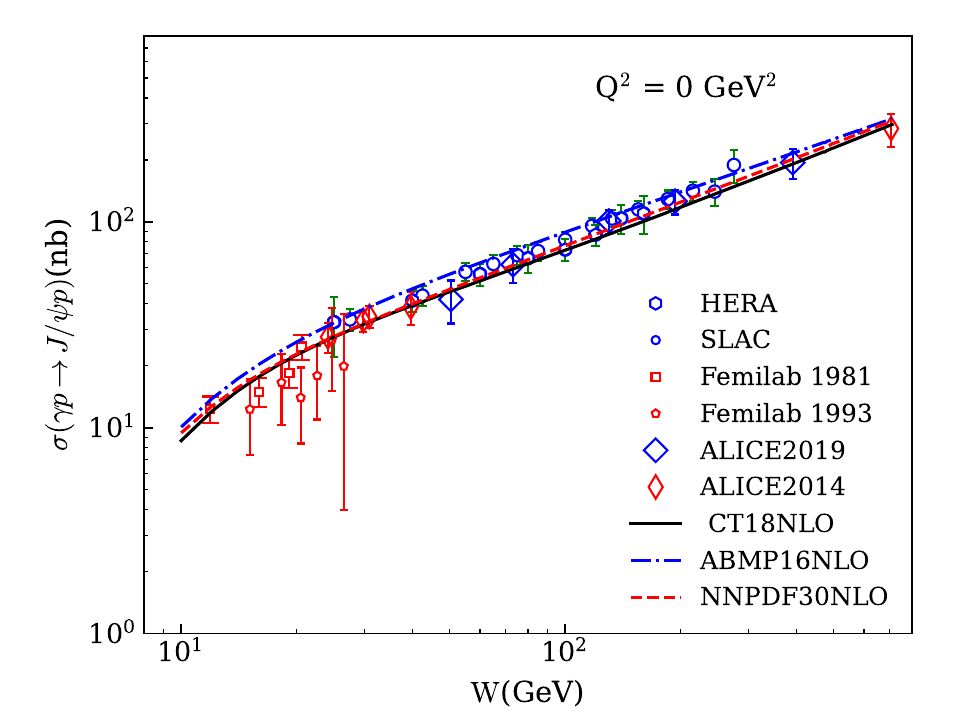}
	\caption{$J/\psi$ total cross section
		  vs $Q^2$ at HERA energy W = 90 GeV (left graph) and at $Q^2$ = 0 GeV$^2$ vs W from low to very high energies(right graph).  The experimental data
	  are taken from HERA\cite{EPJC24}, ALICE\cite{ALICE:2014eof,ALICE:2018oyo},SLAC\cite{Camerini:1975cy},FemiLab\cite{Binkley:1981kv,E687:1993hlm}.}
	\label{fig03a}
\end{figure}

Furthermore, we present comparison of our model results on $J/\psi$ production with HERA data at $W$=90 $\mbox{GeV}$. They are shown in Fig. \ref{fig03a}, left graph is total
cross sections as a function of $Q^2$. Models describe the data well in all $Q^2$ range. In a moment we have photoproduction cross section ($Q^2$ = 0 $\mbox{GeV}^2 $ ) for very high energies from ALICE experiments at LHC \cite{ALICE:2014eof,ALICE:2018oyo}. In Fig.~\ref{fig03a}, right graph shows cross section from low $W \sim 10 $GeV till high $W \sim 700 $GeV energies. It can be seen that all GPDs models give a good description to experimental data in a wide energy range. Note that there are similar data from LHCb \cite{lhcb}, that are compatible with ALICE date. We do not show them on the graph to do data more distinguishable.

In addition, we shall discuss the spin observables, see (Section III): the transverse
target single spin asymmetry (SSA) $A_N$ and double spin asymmetries (DSAs)
$A_{LS}$ that requiring a polarized target and are connected with $E_g$  GPDs.
The $E_g$ GPDs are determine proton spin-flip amplitude Eq.~(6) which is similar
in form with non-flip amplitude Eq.~(5) with exception of the kinematical
prefactor. Effects of gluon  $E_g$ can be observed in $A_N$ asymmetry and DSAS
$A_{LS}$. Previous analyses of these effects in EHMP can be found in the Ref.
\cite{Koempel:2011rc}.

\begin{figure}[!h]
	\centering
	\includegraphics[width=7.5cm]{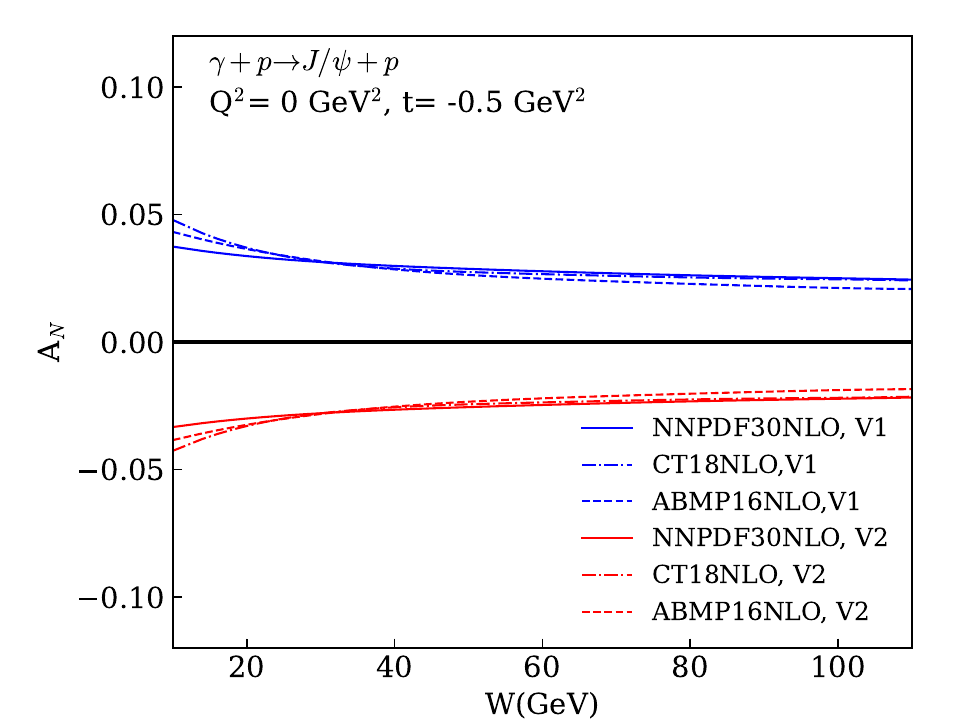}
	\includegraphics[width=7.5cm]{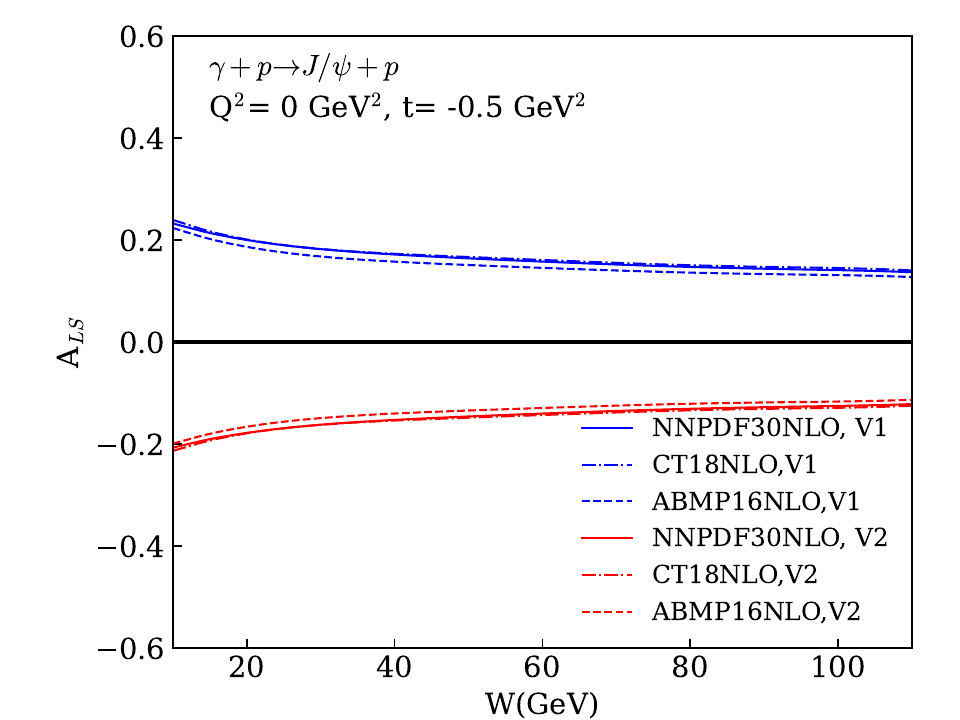}
	\caption{$J/\psi$ single spin asymmetry $A_N$ and double spin asymmetry $A_{LS}$ versus W at $Q^2$ = 0 GeV$^2$. }
	\label{fig04}
\end{figure}

 To estimate $E_g$ we use parameterizations (2,3) from \cite{gk08},that called in the
 graphs as (V1,V2), see Fig~\ref{fig04}. Energy dependencies of this asymmetries
 are determined  by low-$x$ dependence of $x e_g(x)$ PDF $\sim 1/x^\lambda$ that is similar to $xh_g(x)$ behavior at
 low-$x$ range. As a consequence, $W$ dependencies of these asymmetries are rather
 weak. Our results on $A_N$ and $A_{LS}$ asymmetries are shown in
 Fig.~\ref{fig04} for $J/\psi$ photoproduction. We find that  $A_N \sim 0.05$
 and $A_{LS} \sim 0.2$. These results are not far from estimations in
 \cite{Koempel:2011rc}, where more variants of $E_g$ parametrization are
 presented. We can conclude that these $J/\psi$ asymmetries can be measured in
 both future EicC and EIC-US colliders. As a result important information
 on gluon $E^g$ GPDs can be extracted. With the help of Ji sum rules \cite{ji1},
 it permit us to estimate the gluon angular momentum in the hadron referring the discussion
 in \cite{Koempel:2011rc}.

The $\tilde {H}^g$ contribution to $M_{++,++}$ amplitude is essential in $A_{LL}$
asymmetry. It is determined by the second term of Eq.~(5). The corresponding $f^-
_{11}$ function is shown in Eq.~(12). It is proportional to $k_{\perp}^2$ and
suppressed by the factor $\langle k_{\perp}^2\rangle /m_V^2$ with respect to $f^+_{11}$ function of
Eq.~(11) that is accompanied by $H_g$ contribution to the amplitude.  Effects of
$\tilde{H}_g$ can be tested in $A_{LL}$ asymmetry. Corresponding results for
these asymmetries in light vector meson $\rho$, $\phi$ production can be found in \cite{gk06, gk07}.
	
\begin{figure}[!h]
	\centering
	\includegraphics[width=7.5cm]{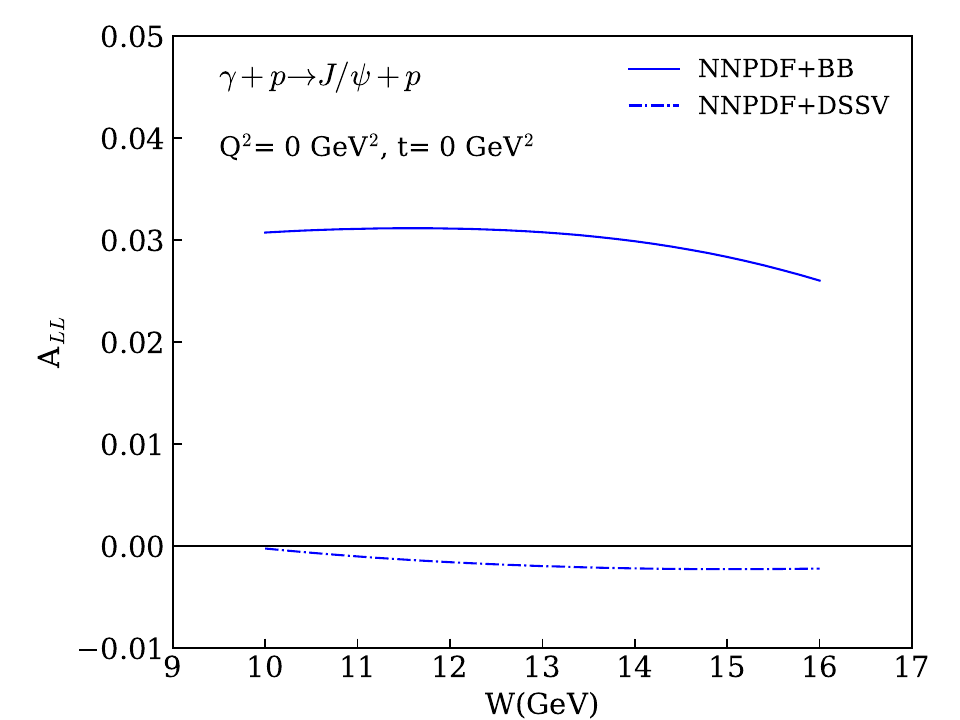}
	\includegraphics[width=7.5cm]{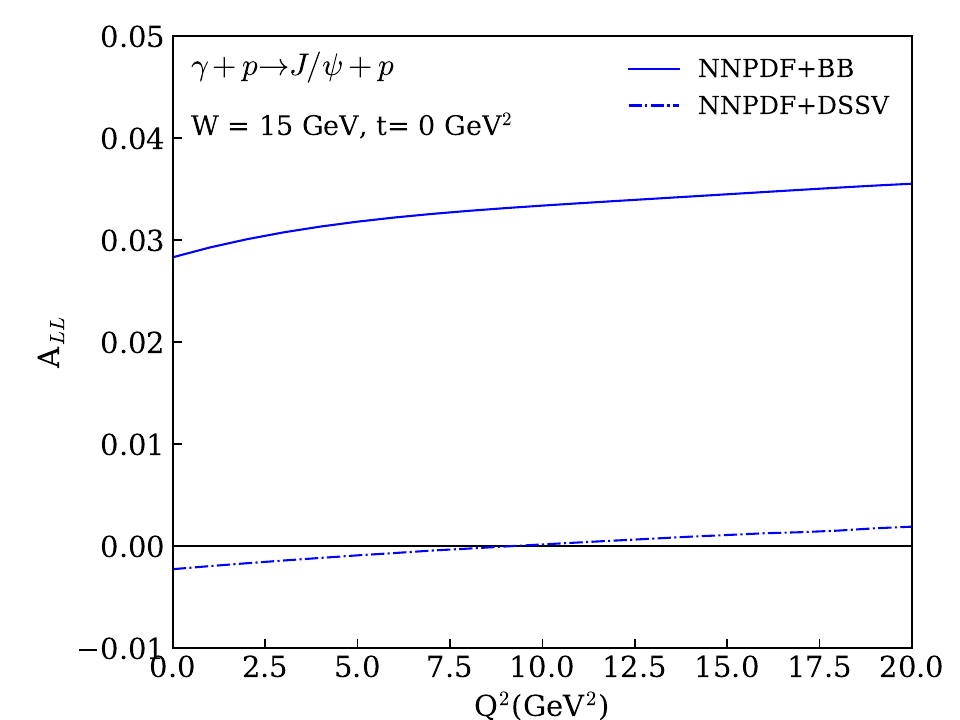}
	\caption{Helicity correlation $A_{LL}$ of $J/\psi$ at different Q$^2$ and W . }
	\label{fig05}
\end{figure}

To study $A_{LL}$ asymmetry we use BB \cite{BB} and DSSV
\cite{DSSV} parameterizations of $\Delta g$ PDFs which determine $\tilde{H}^g$
GPDs through  DD representation  Eq.~(18). $A_{LL}$ asymmetry decreases with energy
quite rapidly and at EIC-US energies it is close to zero. Hence, we present in
Fig.~\ref{fig05} the $W$ and $Q^2$ dependencies of $A_{LL}$  of $J/\psi$ asymmetry
at energies, typical for EicC. We find that $A_{LL} \sim 0.03$ for BB case and
$A_{LL} \sim 0.0$ for DSSV parameterizations. Thus, study of this asymmetry at
EicC can give an answer about of value of $\tilde{H}^g$ convolution functions.

\section{Conclusion }
In this paper we analyse the heavy vector meson electroproduction
within the handbag approach. Because of the large scale determined
by meson mass $m_V \sim$ 3 GeV production amplitude even at
$Q^2=0$ GeV$^2$ factorizes  into the hard subprocess  and GPDs.  The
hard subprocess amplitude  was calculated perturbatively,
 with taken into account the transverse quark momenta.
It was mentined that the Sudakov factors are not important here because
of the nonrelativistic heavy meson wave function used in hard subprocess.
We calculate non-flip amplitudes for longitudinal and transverse photon polarization
determined by GPDs $H_g$, proton spin-flip amplitude containing GPD $E_g$. In addition
contribution determined by  $ \tilde H_g$ GPDs is calculated. It is suppressed by
the factor $\langle k_\perp^2\rangle /m_V^2$ with respect to GPDs $H_g$ conribution.

In our approach we parameterize gluon GPDs using three sets of PDFs  with the
help of double distribution representation. With these  PDFs, we perform $J/\psi$
differential cross sections and total cross sections in GPD approach. The
results of our model are compared with experimental data. It can be
found that the our predictions  can give a good agreement  with the HERA H1 and ZEUS
$J/\psi$ cross sections data. It seems that CT18NLO gluon  PDF gives a little bit better
results.

In GPD approach using DD representation with corresponding PDFs we estimate spin asymmetry observables.
For two sets of GPDs we calculate single spin $A_N$ and double spin $A_{LS}$ asymmetries
determined by $E_g$ GPDs. These asymmetries have a weak energy dependence and can be measured
at low EicC and at  high Eic energies. Test of this contribution is important to determine
gluon contribution to proton angular momentum $J_g^P$ with the help of Ji sum rules \cite{ji1}. More
discussion on this issue can be found in \cite{Koempel:2011rc}.

We calculate $A_{LL}$ asymmetry determined by  $\tilde{H}_g$ GPDs.  This asymmetry decreases
with energy very rapidly and can be measured only at low EicC energies. Test of this asymmetry
can give constraints on $\tilde{H}_g$ GPDs.

 EicC can be a good platform to measure single and double spin asymmyetries.
 It can help us to extract the property of GPDs via spin asymmetries measurements.
  We can conclude that the heavy vector meson electroproduction is a good observables to probe various GPDs.

\section*{Acknowledgment}
S.G. expresses his gratitude to P.Kroll for long time collaboration.
This work is partially supported by the NFSC grant (Grant No 12293061) and CAS president's international fellowship initiative (Grant No. 2021VMA0005)
and Strategic Priority Research Program of Chinese Academy of Sciences (Grant NO. XDB34030301) .

\end{document}